\documentclass[aps,english,prb,reprint,superscriptaddress, 
citeautoscript,showpacs]{revtex4-1}
\usepackage{graphicx}
\usepackage{amssymb}
\usepackage{babel}
\usepackage{color}
\usepackage{multirow}
\usepackage{setspace}
\usepackage{longtable}
\usepackage{amsmath}
\usepackage{amsfonts}
\usepackage{amssymb}
\usepackage{booktabs}
\usepackage{hyperref}
\usepackage{xcolor}
\usepackage{braket}
\usepackage{blkarray}
\begin{document}

%\title{A First-principles study of MoSi$_{2}$N$_{4}$ nanoribbons} 
%\title{Prediction of magnetism and Dirac-semimetal character in the MoSi$_{2}$N$_{4}$ nanoribbon by first principles calculations}
%Magnetic and electronic structure of Dirac semimetal in the MoSi$_{2}$N$_{4}$ nanoribbon by first principles calculations
\title{Band-gap engineering, magnetic behavior and Dirac-semimetal character in the MoSi$_{2}$N$_{4}$ nanoribbon with armchair and zigzag edges}

\author{A. Bafekry}\email{Asadollah.Bafekry@uantwerpen.be}
\affiliation{Department of Physics, University of Antwerp, Groenenborgerlaan 171, B-2020 Antwerp, Belgium}
\author{M. Faraji}
\affiliation{Micro and Nanotechnology Graduate Program, TOBB University of Economics and Technology, Sogutozu Caddesi No 43 Sogutozu, 06560, Ankara, Turkey}
\author{C. Stampfl}
\affiliation{School of Physics, The University of Sydney, New South Wales 2006, Australia}
\author{I. Abdolhosseini Sarsari}\email{abdolhosseini@iut.ac.ir}
\affiliation{Department of Physics, Isfahan University of Technology, Isfahan, 84156-83111, Iran}
\author{A. Abdollahzadeh Ziabari}
\affiliation{Nano Research Lab, Lahijan Branch, Islamic Azad University, Lahijan, Iran}
\author{N.~N.~Hieu}
\affiliation{Institute of Research and Development, Duy Tan University, Da Nang 550000, Viet Nam}
\affiliation{Faculty of Natural Sciences, Duy Tan University, Da Nang 550000, Viet Nam}
\author{S. Karbasizadeh}
\affiliation{Department of Physics, Isfahan University of Technology, Isfahan, 84156-83111, Iran}
%\author{M. Yagmurcukardes}
%\affiliation{Department of Physics, University of Antwerp, Groenenborgerlaan 171, B-2020 Antwerp, Belgium}
\author{M. Ghergherehchi}
\affiliation{College of Electronic and Electrical Engineering, Sungkyunkwan University, Suwon, Korea}

\begin{abstract} 
Motivated by the recent successful formation of the MoSi$_{2}$N$_{4}$ monolayer [Hong {\em et al.}, Sci. 369, 670 (2020)], the structural, electronic and magnetic properties of MoSi$_{2}$N$_{4}$ nanoribbons (NRs) is investigated for the first time . The band structure calculations showed spin-polarization in zigzag edges and a non-magnetic semiconducting character in armchair edges. 
For armchair-edges, we identify an indirect to direct band gap shift compared to the MoSi$_{2}$N$_{4}$ monolayer, and its energy gap increases with increasing NR width. Anisotropic electrical and magnetic behavior is observed via band structure calculations in the zigzag and armchair edges, where, surprisingly, for the one type of zigzag-edges configuration, we identify a Dirac-semimetal character. 
The appearance of magnetism and Dirac-semimetal in MoSi$_{2}$N$_{4}$ ribbon can give rise to novel physical properties, which could be useful in applications for next-generation electronic devices.
\end{abstract}
\maketitle

%\section{Introduction}
The intense research into the properties and applications of graphene \cite{1} has triggered an enormous effort to discover novel two-dimensional (2D) layered nanomaterials with peculiar physical properties leadint to new performances. For example, a group of 2D transition metal dichalcogenides (TMDCs), including MoS$_2$ and WS$_2$, has significantly influenced the world of optoelectronic, photonic and energy devices over the last few  years \cite{2,3,4,5,6}. In addition to 2D TMDCs, nanodimensional transition metal nitride (TMN) compounds have been intensively investigated recently \cite{7,8,9,10,11,12}.  
As a new class of air-stable 2D TMN semiconductors, the MoSi$_2$N$_4$ monolayer exhibits unique electrical and mechanical properties. Recently, Ren {\it et al.} have managed to synthesise an air-stable MoSi$_2$N$_4$ 2D monolayer using chemical vapor deposition (CVD) \cite{13}. 
The experimental  bandgap reported in Ref. \cite{14} was 1.94 eV, very close to the value obtained by density functional theory (DFT) calculations.
Moreover, the  intrinsi electron and hole mobilities at the K point in the Brillouin zone were evaluated to be $\sim$270 and $\sim$1200 cm$^{2}$V$^{-1}$s$^{-1}$, respectively; greater than those of monolayer MoS$_2$ by factors of 4 to 6 \cite{14}. Furthermore, a high optical transmittance was obtained, with an average of 97.5 $\pm$ 0.2\% in the visible range, comparable to that of 0.335-nm-thick monolayer graphene (97.7$\%$). These promising aspects make the MoSi$_2$N$_4$ monolayer  an appropriate candidate for use in nano-electronic and optoelectronic devices.

These attractive properties pave the way not only to explore novel 2D devices but also to investigate the structural, electrical, mechanical and optical properties of MoSi$_2$N$_4$. To date, the physics behind electrically contacting MoSi$_2$N$_4$ with metals is still unknown \cite{15,16,17}. 
Nanoribbons are narrow strips of a material with a very high length-to-width ratio.They combine the flexibility and unidirectional properties of one-dimensional nanomaterials, the high surface area of 2D nanomaterials and the electron-confinement and edge effects of both. The structures of nanoribbons can thus lead to exceptional control over electronic band structure, the emergence of novel phenomena and unique architectures for applications \cite{18,19,20,21,22,23,24}. As the flagship of nanoribbons, graphene nanoribbons (GNRs) were initially introduced as a theoretical model to examine the edge and nanoscale size effects in graphene \cite{25}. The electronic states of GNRs largely depend on the edge structures (armchair or zigzag). Meanwhile, black phosphorous has become a strong focus of the science community owing to the synthesis of its 2D form, namely, the phosphorene \cite{26,27,28}. Preliminary, but exciting results indicate that phosphorene has potential applications in nanoelectronics and optoelectronics. It has been found that all the semiconducting phosphorene nanoribbons present very high values of Seebeck coefficient and can be further enhanced by hydrogen passivation at the edge \cite{29}. There are also reports on nanoribbon structures of different materials such as silicene  \cite{30}, boron nitride \cite{31}, gallium oxide \cite{32} and titanate nanoribbons \cite{33}. 

Motivated by the exciting experimental realization of 2D MoSi$_2$N$_4$, in this study our objective is to explore the structural, electronic and magnetic properties of MoSi$_2$N$_4$ nanoribbons using {\em ab initio} calculations and to investigate the effect on these properties of changing the width of zigzag and armchair nanoribbons. 
Our results show MoSi$_2$N$_4$ nanoribbons with armchair edge possess a semiconducting behavior and the band gap increases with the width of the NR, while MoSi$_2$N$_4$ nanoribbons with zigzag edges show magnetic metal properties, where the magnetic moment is modified with the width of the NR.
Remarkably, half-metallic anisotropic Dirac cone is found in one of configurations of MoSi$_2$N$_4$ nanoribbons which makes them worth systematically investigating.

\section{Method}
The plane-wave basis projector augmented wave, as implemented in the Vienna {\em ab-initio} Simulation Package (VASP) \cite{vasp1,vasp2} was employed in the framework of DFT. The generalized gradient approximation (GGA) of Perdew-Burke-Ernzerhof form, \cite{GGA-PBE1,GGA-PBE2} were used for the exchange-correlation functional. The kinetic energy cut-off of was 600 eV, and a $\Gamma$-centered 16$\times$16$\times$1 {\it k}-mesh for the unit cell was employed in our calculations. The tolerance of the total energy convergence was less than 10$^{-5}$ eV with forces less than 10$^{-3}$ eV \AA{}$^{-1}$. The lattice constants and atomic positions were optimized without any constraint. The vacuum space was $\sim$20 \AA{} along the \textit{z}-direction to avoid any fictitious interactions. 

\section{Monolayer}
%\section{Structure and Stability}

\begin{table*}
\centering
\caption{\label{table1} The structural and electronic parameters including lattice constant $a$;
the bond lengths of Mo-N ($d_{1}$), Si-N ($d_{2}$) and Mo-N ($d_{3}$);
the bond angles of N-Mo-N ($\theta_{1}$), N-Si-N ($\theta_{1}$) and N-Mo-N ($\theta_{1}$);
%the bucking of MoSi$_{2}$N$_{4}$ defined by the difference between the largest and smallest z coordinates of Zn and Sb atoms $(\Delta{z})$;
the layer thickness   $(t)$;
the cohesive energy per atom, $(E_{coh})$;
%the charge transfer $(\Delta{Q})$ between atoms;
the work function ($\Phi$);
the band gap $(E_{g})$ using PBE (HSE06); the VBM/CBM positions.
}
\begin{tabular}{lccccccccccccc}
\hline\hline
&\textit{a} & \textit{d$_{1}$}& \textit{d$_{2}$}& \textit{d$_{3}$} &$t$ & \textit{$\theta_{1}$}&\textit{$\theta_{2}$}&\textit{$\theta_{3}$}& $E_{coh}$ & $\Phi$ &$E_{g}$& VBM/CBM\\
& (\AA) & (\AA) & (\AA) & (\AA) & (\AA) &($^{\circ}$) & ($^{\circ}$) & ($^{\circ}$) & (eV/atom) & (eV) & (eV) & (eV) & \\
\hline
MoSi$_{2}$N$_{4}$            & 2.91  & -    & 1.75 & 2.09 & 7.01  & -     & 112.02 & 73,87 & -8.46 & 5.12 & 1.79 (2.35) & $\Gamma$/K\\
\hline\hline
\end{tabular}
\end{table*}

\begin{figure}[!b]
\includegraphics[scale=1]{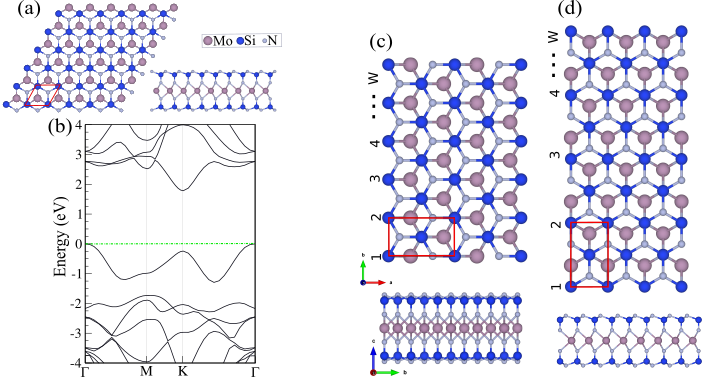}
\caption{(a) The geometry (top and side views) and (b) band structure of the 2D MoSi$_{2}$N$_{4}$ nanosheet. Geometric structures of the (c) Armchair and (d) Zigzag MoSi$_{2}$N$_{4}$ NRs with different  widths specified according to the numbers. The primitive unit cell is indicated by a red parallelogram (rectangle) for the nanosheet (nanoribbon).}
\label{1}
\end{figure}

The atomic and electronic structures of the MoSi$_{2}$N$_{4}$ nanosheet are shown in Figs. \ref{1}(a) and (b), respectively.
From the atomic structure, we can see that atoms are packed in a honeycomb lattice, forming a 2D crystal with a space group of $P_{6m1}$. The hexagonal primitive unit cell is indicated by the red parallelogram.
After structure optimization the lattice constant is determined to be 2.91 \AA{}.
The bond lengths of Mo-N and Si-N are determined to be 1.75 and 1.74 \AA{}, respectively, while the bond length of Mo-Si is calculated to be 2.09 \AA{}. The two angles of Si-N-Si are determined to be 112 and 106$^{\circ}$, while the two angles of Si-Mo-Si are 87 and 73$^{\circ}$. 
The thickness of MoSi$_{2}$N$_{4}$ is 7.01 \AA{} and for the MoN$_{2}$, the vertical distance of N-N is determined to be 2.99 \AA{} and the Mo-N bond length is 2.58 \AA{}. These results are good agreement with a previous report \cite{13}.
 
The cohesive energy ($E_{coh}$) per atom is calculated using the following equation:
\begin{equation}
E_{coh} = \frac{E_{tot} - E_{Mo}- 2E_{Si} - 4E_{N}}{n_{tot}},
\end{equation}
where $E_{Mo}$, $E_{Si}$ and $E_{N}$, $E_{tot}$ represent the energies of isolated Mo, Si, N atoms and total energy of the ML, n$_{tot}$ is the total number of atoms, respectively. 
The cohesive energy is found to be -8.46 eV/atom for MoSi$_{2}$N$_{4}$, while the negative energy indicates the exothermicity of the structure. 
The electronic band structure of MoSi$_{2}$N$_{4}$ is shown in Fig. \ref{1}(b). 
Our results show that MoSi$_{2}$N$_{4}$ is semiconductor with an indirect band gaps of 1.79 eV (2.35), using the PBE (HSE06)~\cite{Heyd} functional, where the VBM and CBM are located at the $\Gamma$ and the K-point, respectively. 

\begin{figure*}[!htb]
	\includegraphics[scale=1.2]{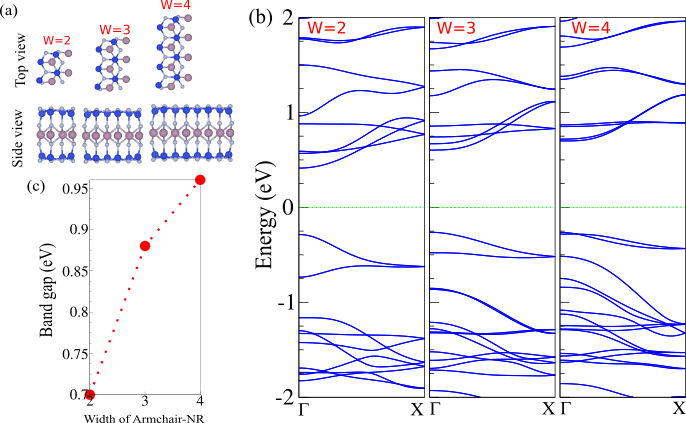}
	\caption{(a) Optimized structures, (b) electronic band structure of armchair MoSi$_{2}$N$_{4}$ NR with different widths. (c) Variation of the band gap with respect to the different widths.}
	\label{2}
\end{figure*}

\begin{figure}[!b]
	\includegraphics[scale=1]{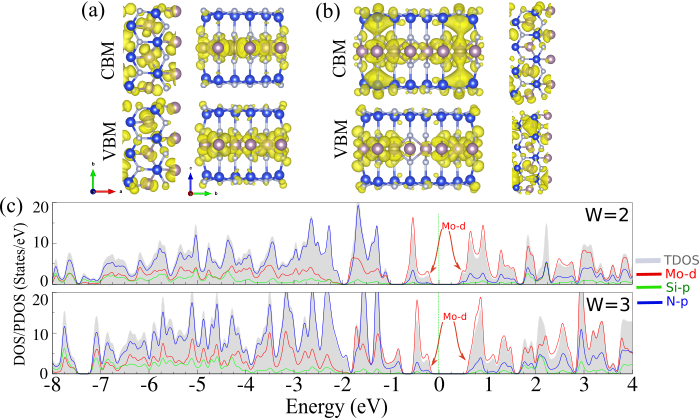}
	\caption{Charge density of valence band maximum (VBM) and conduction band minimum (CBM) of Armchair MoSi$_{2}$N$_{4}$ NR for the (a) $W=2$ and (b) $W=3$ widths (top and side views). (c) Density of states (DOS) and projected DOS (PDOS).}
	\label{3}
\end{figure}

\section{Armchair edge Nanoribbon}
%armchair

As shown in Fig.~\ref{1}(c) and (d), the structure of the MoSi$_{2}$N$_{4}$ NRs is specified with different widths. In the monolayer nanosheet, all atoms have the same atomic environment which make their bands degenerate. However, in the NRs, the existence of different environments for different rows of atoms changes the degeneracy of the bands of the respective atoms, resulting in the presence of numerous non-degenerate energy bands in the band structure. 

The optimized atomic structure and electronic band structure for different widths in the armchair edges are shown in Figs.~\ref{2}(b-c), respectively. 
Notice that the NRs are direct bandgap semiconductors, and we can see that the bandgap with increasing increases with increasing width. 
Compared to the monolayer nanosheet, formation of NRs create a change from an indirect semiconductor behavior to direct gap. 
Focusing on the band structures of MoSi$_{2}$N$_{4}$ NRs as illustrated in Fig.~\ref{2}(b), it is found that both the valence band maximum (VBM) and the conduction band minimum (CBM) locate at the $\Gamma$-point in the Brillouin zone. Interestingly, the subbands near the Fermi level get flatter as the width of NR increases. The phenomenon that energy subbands near the Fermi level become flatter for wider nanoribbon has also been observed in armchair graphene nanoribbons~\cite{Fujita1996}.

Also, as depicted in Fig.~\ref{2}(c), the energy gap of MoSi$_{2}$N$_{4}$ NR tends to increase more slowly over wider ribbon. 
The calculated band gaps are 0.7 eV (W=2), 0.88 eV (W=3) and 0.96 eV (W=4)

In Fig.~\ref{3} we show the density of states (DOS) and the electron density distribution of the VBM and the CBM for two armchair NRs. Fig.~\ref{3}(c) shows spurious Mo-$d$/N-$p$ hybridization which because of $p$-$d$ repulsion pushes the N-$p$ states higher in energy reducing $\Delta E$. Since the Mo-$d$/N-$p$ hybridization in armchair NR with $W=2$ is more than  armchair NR with $W=3$, energy reducing $\Delta E$ compete with effects of quantum confinement. This effect, which can be traced to the self-interaction error, is over and above the conventional DFT underestimation of band gaps in semiconductors\cite{Fuchs2007}.

The charge density of the VBM is concentrated inside the armchair MoSi$_{2}$N$_{4}$ NRs, while the charge density of the CBM distributes towards the edges with increasing width. This property suggests the potential for tailoring and engineering of the transport by tuning the width. In the monolayer nanosheet, the CBM and the VBM are located at the K-point and $\Gamma$-point, respectively, while in the armchair MoSi$_{2}$N$_{4}$ NRs, both CBM and VBM are located at the $\Gamma$-point. This transition from indirect to direct bandgap is the effective consequence of quantum confinement along the $\Gamma$--K path in armchair MoSi$_{2}$N$_{4}$ NRs. 
Regarding to the contribution of the atomic orbitals to the formation of electronic bands of MoSi$_{2}$N$_{4}$ NRs, the image of the partial density of states (PDOS) in Fog.~\ref{3}(c) demonstrate that the Mo-$d$ orbitals make a dominant contribution to the formation of electronic band, especially the conduction band as well as the valence band close to the fermi level. In the both cases of  MoSi$_{2}$N$_{4}$ NRs as depicted in Fig.~\ref{3}(c), the valence band is formed from the main contributions of the Mo-$d$, Si-$p$, and N-$p$ orbitals. In particular, except in the vicinity of the Fermi level, from $-0.2$ to $-1.2$~eV, the N-$p$ orbitals's contribution to the valence band is even greater than that of other orbitals, including Mo-$d$ orbitals.

\section{Zigzag edge nanoribbon}

\begin{figure}[!b]
	\includegraphics[scale=1]{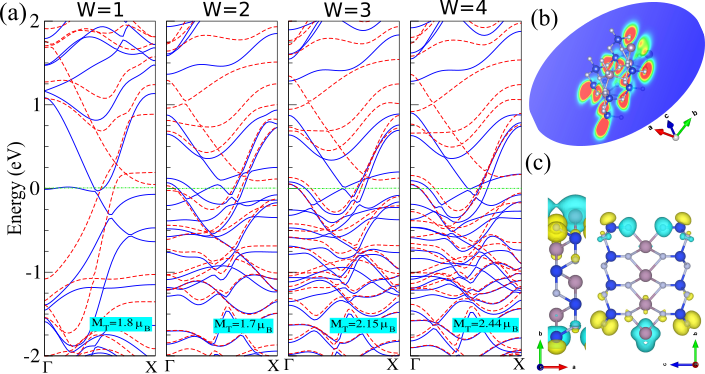}
	\caption{(a) Electronic band structure of zigzag MoSi$_{2}$N$_{4}$ NR for different widths. (b) Contour plot of the electron localization function (ELF) of zigzag NR for w=2. Red (blue) color indicates high (low) electron density. (c) Difference spin density of zigzag NR for w=2. The green and yellow regions represent the $\uparrow$ and $\downarrow$ spin states, respectively. The zero of energy is set to Fermi-level.}
	\label{4}
\end{figure}

The scenario for zigzag MoSi$_{2}$N$_{4}$ NRs is totally different to the armchair NRs. The band structure calculations in Fig.~\ref{4}(a) shows metallic behavior for zigzag NRs with sufficient states at the Fermi surface to satisfy the Stoner criterion, thus the zigzag NRs exhibit magnetization. Total magnetizations for zigzag NRs with widths 1, 2, 3, and 4 are 1.8, 1.7, 2.15, and 2.44 $\mu_B$, respectively. 
The zigzag NRs with odd and even widths experience different magnetic symmetries so that we should compare odd and even widths, separately. For both groups, we can observe the total magnetization increases with the width, where the spin density occurs at the edges of the zigzag NRs (see Fig.~\ref{4}b-c). With increasing   width of the zigzag NRs, the region of spin polarization penetrates inside the zigzag NRs which results in higher magnetization. 
The anisotropy observed in the different MoSi$_{2}$N$_{4}$ NRs, that is, the spin-polarized magnetic zigzag NRs versus the non-magnetic semiconductor armchair NRs, is attractive for spintronic applications.
Difference from armchair NRs, the band structures of the zigzag NRs show a metallic character for both spin configurations as depicted in Fig.~\ref{4}(a). 
The band structures of zigzag NRs are asymmetric for all investigated ribbon-widths and spin configurations. There are some energy levels (two or three) in each spin configuration cross the Fermi level and eliminate the band gap of zigzag NRs. Also, metallic characteristic of zigzag NRs is found in all investigated width $W$ of nanoribbon $(W = 1,2,3,4)$.

\begin{figure}[!t]
	\includegraphics[scale=1]{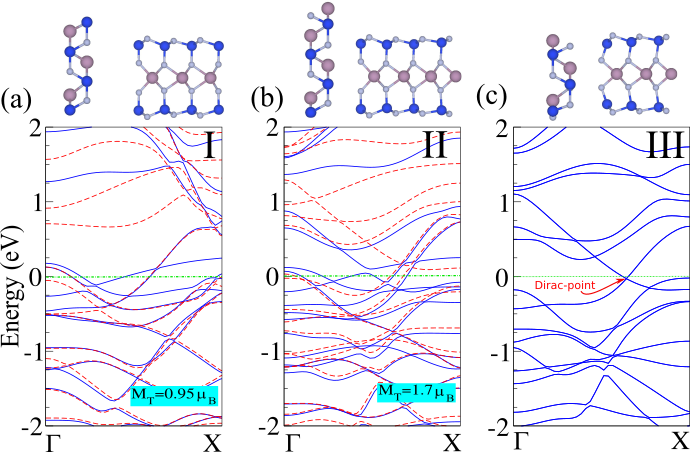}
	\caption{(a) Optimized structures of zigzag MoSi$_{2}$N$_{4}$ NR with respect to (a) I, (b) II and (c) III configurations. Electronic band structure with corresponding each NRs is shown in bottom panel.  The zero of energy is set to Fermi-level.}
	\label{5}
\end{figure}

Investigating more, we specify two configuration, Figs.~\ref{5}(a-c), near the zigzag MoSi$_{2}$N$_{4}$ NRs with $W=2$, Fig.~\ref{4}(b). 
As can be seen, in configuration I, the total magnetization decreases to 0.95 $\mu_B$, while in configuration III, the total magnetization vanishes and a tilted Dirac cone appears which is promising for valley filtering in p-n junctions\cite{Nguyen2018} and the generation of a photocurrents\cite{Chan2017}. 
The synthesis of this kind of Dirac point is challenging \cite{Hirata2016} as they require materials whose constituent atoms are arranged in lattices with intricate electron hopping\cite{Soluyanov2015}. In the configurations I and II, some subbands across the Fermi level and close the energy gap leading to both configurations preserving the metallic behavior as shown in Figs.~\ref{5}(a,b). Interestingly, the Dirac coin is found in the configuration III as illustrated in Fig.~\ref{5}(c). However, unlike for pristine graphene, the Dirac coin for MoSi$_{2}$N$_{4}$ NR as presented in Fig.~\ref{5}(c) is anisotropic because the linear dispersions along $\Gamma$--K and its perpendicular direction are different. This may lead to the anisotropy in the Fermi velocities in the configuration I of MoSi$_{2}$N$_{4}$ NR which is due to the nature of one-dimensional nanoribbon structure. Half-metallic behavior with anisotropic Dirac point was also found in zigzag graphene NRs on graphene~\cite{Chen2016}. This finding could be useful in applications for electronic devices.

\section{Conclusion}
In conclusion, we investigated the electronic properties of MoSi$_{2}$N$_{4}$ nanoribbons using first-principles calculations. 
The electronic properties showed spin-polarization in 
zigzag-edges MoSi$_{2}$N$_{4}$ NRs and semiconducting character in armchair-edges MoSi$_{2}$N$_{4}$ NRs. 
In armchair edges of MoSi$_{2}$N$_{4}$ nanoribbons, we identify an indirect to direct band gap shift compared to the MoSi$_{2}$N$_{4}$ monolayer, where the bandgap increase with increasing width.
%This interesting and anisotropic behavior of the nanoribbons is potentially promising for applications in spintronic devices.
The appearance of the Dirac-semimetal in one type of MoSi$_{2}$N$_{4}$ NRs with zigzag-edges is a wonderful feature, which can give rise to many interesting physical properties, such as quantum electronic transport.
 
\section{Declaration of competing interest}
The authors declare that they have no known competing financial interests or personal relationships that could have appeared to influence the work reported in this paper.

\section*{ACKNOWLEDGMENTS}
This work was supported by the National Research Foundation of Korea (NRF) grant funded by the Korea government (MSIT) (NRF-2015M2B2A4033123).

\section*{DATA AVAILABILITY}
The data that support the findings of this study are available from the corresponding author upon request.

\end{document}